\documentclass[a4paper,12pt]{elsarticle}

\usepackage{lineno,hyperref}
\usepackage{color}                        % For color text: \color
\usepackage{xspace}
\usepackage{pdfwidgets}
\usepackage{amssymb}                    % useful mathematical symbols
\usepackage{pdflscape}
\usepackage{siunitx}
\usepackage{multirow}

\modulolinenumbers[5]

\journal{JASTP}

\biboptions{authoryear}

\newcommand{\ie}{{\it i.e.~}}
\newcommand{\eg}{{\it e.g.}}
\newcommand{\mr}{\mathrm}

\chardef\us=`\_

\begin{document}

\begin{frontmatter}

\title{New estimation of non-thermal electron energetics in the giant solar flare on 28 October 2003 based on Mars Odyssey observations}
%\title{Elsevier \LaTeX\ template\tnoteref{mytitlenote}}





\author[aff1,aff2]{B.A.~Nizamov\corref{mycorrespondingauthor}}
\cortext[mycorrespondingauthor]{Corresponding author}
\ead{nizamov@physics.msu.ru}

\author[aff3,aff4,aff5]{I.V.~Zimovets}
\author[aff3]{D.V.~Golovin}
\author[aff3]{A.B.~Sanin}
\author[aff3]{M.L.~Litvak}
\author[aff3]{V.I.~Tretyakov}
\author[aff3]{I.G.~Mitrofanov}
\author[aff3]{A.S.~Kozyrev}

\address[aff1]{Sternberg Astronomical Institute of Lomonosov Moscow State University, Universitetsky pr. 13, Moscow 119234, Russia}
\address[aff2]{Faculty of Physics, M.V.Lomonosov Moscow State University, Leninskie Gory, Moscow 119991 Russia}
\address[aff3]{Space Research Institute (IKI) of the Russian Academy of Sciences, Profsoyuznaya str. 84/32, Moscow 117997, Russia }
\address[aff4]{State Key Laboratory of Space Weather, National Space Science Center (NSSC) of the Chinese Academy of Sciences, No.1 Nanertiao, Zhongguancun, Haidian District, Beijing 100190, China  }
\address[aff5]{International Space Science Institute -- Beijing (ISSI-BJ), No.1 Nanertiao, Zhongguancun, Haidian District, Beijing 100190, China }

\begin{abstract}
A new estimation of the total number and energy of the non-thermal electrons produced in the giant ($>\text{X}17$) solar flare on 2003 October 28 is presented based on the analysis of the observations of the hard X-ray (HXR) emission by the High Energy Neutron Detector (HEND) onboard the Mars Odyssey spacecraft orbiting Mars. Previous estimations of the non-thermal electron energy based on the Reuven Ramaty High-Energy Solar Spectroscopic Imager (RHESSI) data were incomplete since RHESSI missed the peak of the flare impulsive phase. In contrast, HEND observed the whole flare. We used two models to estimate the energy of the non-thermal electrons: the cold thick target model and the warm thick target model. More specifically, in the second case we employed an approximation which relates the pitch-angle averaged injection spectrum with the electron spectrum integrated over the emitting source. We found that, depending on the model used and the low-energy cutoff ($E_\mr{c}$) of the non-thermal electrons, the estimate of their total energy in the entire flare can vary from $2.3 \times 10^{32}$ to $6.2 \times 10^{33}$ ergs. The lowest estimate, $2.3 \times 10^{32}$ ergs, obtained within the cold thick target model and fixed $E_\mr{c}=43$ keV, is consistent with the previous estimate. In this case, non-thermal electrons accelerated in the peak of the flare impulsive phase missed by RHESSI contained approximately $40\%$ of the total energy of non-thermal electrons of the entire flare. The highest value, $6.2 \times 10^{33}$ ergs, obtained with the cold thick target model and fixed $E_\mr{c}=10$ keV, looks abnormally high, since it exceeds the total non-potential magnetic energy of the parent active region and the total bolometric energy radiated in the flare. Our estimates also show that the total number and energetics of the HXR-producing electrons in the flare region is a few orders of magnitude higher than of the population of energetic electrons injected into interplanetary space. 
\end{abstract}
%%%%%


\end{frontmatter}

%\linenumbers

%%%%%%%%%%%%%%%%%%%%%%%%%%%%%%%%%%%%%%%%%%%%%%%%%%%
\section{Introduction}\label{S-Intro} 
Major solar flares have always been an object of intensive investigation. Apart from the fact that they demonstrate a huge variety of physical processes and interactions between them, there are two reasons which make such events worth considering. First, strong solar flares can severely impact the Earth causing geomagnetic storms and affecting technical facilities including spacecraft, airplanes and ground infrastructure. A good illustration of this point is the so-called Halloween storm which was caused by a series of strong solar flares in October-November 2003. A special issue of the Journal of Geophysical Research (volume 110, issue A9) was dedicated to this phenomenon (see, \eg, \cite{Gopalswamy2005} for an overview). Also, the ''Solar Extreme Events-2003'' collaboration organized in Russia presented the detailed investigation of these extreme phenomena in a special volume of the Cosmic Research journal (see \citealt{Veselovsky04,Panasyuk04}).

Second, large solar flares serve as natural benchmarks indicating the highest energy which can be released in such an event. This is interesting from the point of view of solar-stellar relation. \cite{Maehara2012} and \cite{Shibayama2013} report on the so-called superflares on G-type dwarfs. Their estimations of the total energy released in a superflare reach $10^{36}$ ergs. The stars they analyzed are close to the Sun in their fundamental parameters, but they are mostly very young and fast rotating objects. A natural question is whether the difference in energies between solar and stellar flares is only quantitative or also qualitative. To understand this, it is important to understand what the energy limit for solar flares is.

In the last three solar cycles, there occurred a number of exceptional, or "giant" flares. By "giant" one means extremely powerful flares, during which X-ray detectors onboard the Geostationary Operational Environmental Satellites (GOES) are in the saturated state (\eg, \citealp{Kane1995,Struminsky2013}). In the $22^{\text{nd}}$ solar cycle, the level of the GOES saturation corresponded to an X12 class flare. A series of giant flares greater than X12 took place on 1, 4, 6, 11 and 15 on June 1991. During the next ($23^{\text{rd}}$) cycle, when the level of the GOES saturation was increased to around X17, there took place a series of powerful flares in October-November 2003 mentioned above. 

\cite{Kane2005} observed the most extreme event on November 4 with \textit{Ulysses} and estimated the total energetics of non-thermal (\(>20\) keV) electrons produced in the flare to be \(\approx 1.3 \times 10^{34}\) ergs, a value much higher than in any other solar flare ever observed (see also \citealt{Kane1995}). \cite{2005GeoRL..32.3S02B}, \cite{Klassen2005}, \cite{Miroshnichenko2005} and \cite{Simnett05} discuss the timings of energetic particles in, probably, the second largest of the October-Novemeber 2003 events, \ie the October 28 event, but did not estimate the total energetics of non-thermal electrons accelerated in the flare region. \cite{Mewaldt05} estimated the total energetics of non-thermal interplanetary particles (including electrons, protons and ions) to be \(\approx 6 \times 10^{31}\) ergs. \cite{2005IJMPA..20.6705K}, \cite{Grechnev2005} and \cite{2010CosRe..48...70K} observed this event with CORONAS-F. \cite{Kopp2005} detected this flare with the Total Irradiance Monitor. \cite{2004ESASP.552..669G} and \cite{Kiener2006} investigated gamma radiation of this flare by the gamma-ray spectrometer SPI/INTEGRAL. \cite{Su2006} compared the EUV observations of the event made by TRACE and hard X-ray (HXR) observations by the Anti-Coincidence System ACS/INTEGRAL. \cite{Struminsky2013} compared electromagnetic emissions of this and other giant flares of the $23^{\text{rd}}$ solar cycle. It is shown that the peak fluxes of HXR and microwave emissions of the October 28 flare were even higher than in the November 4 flare. This indicates that the energetics of non-thermal electrons in the first flare (having apparently the lower X-ray class) could be even higher than in the second one. Nonetheless, the energetics of electrons accelerated in the October 28 flare were not estimated in the aforementioned papers. 

\begin{figure} 
\centerline{\includegraphics[width=0.8\textwidth]{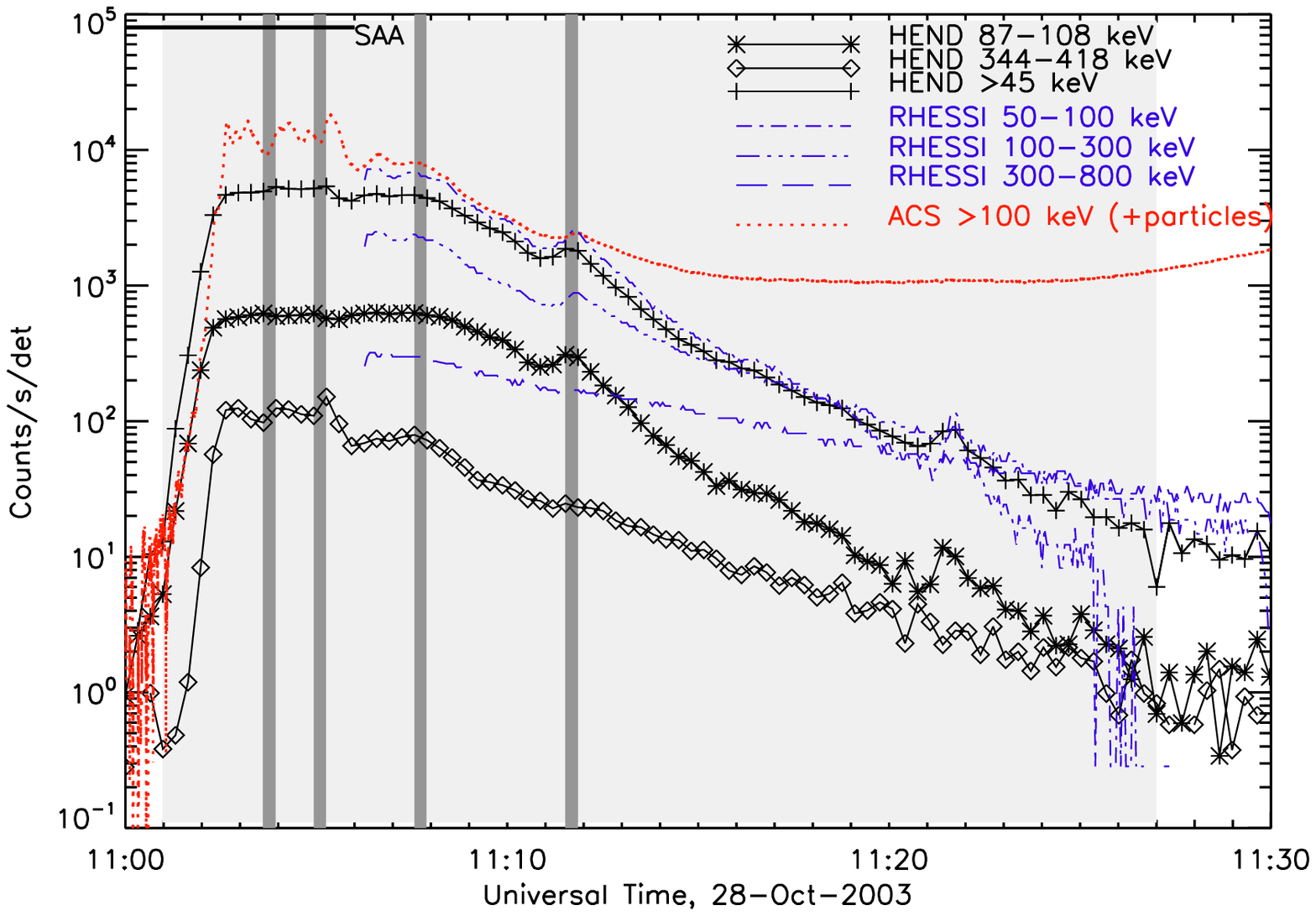}}

\caption{Background-subtracted count rates in different channels of HEND (20 s resolution), RHESSI (4 s resolution) and ACS/INTEGRAL (50 ms resolution, smoothed over 1 s, and divided by 20) during the flare on 28 October 2003. The difference in time on Mars and Earth ($205.46$ s) is taken into account. The time interval when RHESSI was in the SAA is shown at the top. The light gray region indicates the time interval over which the total number and energetics of non-thermal electrons are calculated. Four dark gray vertical stripes indicate the time intervals for which HXR spectra are shown in Fig.~\ref{fig:demo_spec}. } 
\label{fig:countrates}
\end{figure}

\begin{figure} 
\centerline{\includegraphics[width=0.6\textwidth]{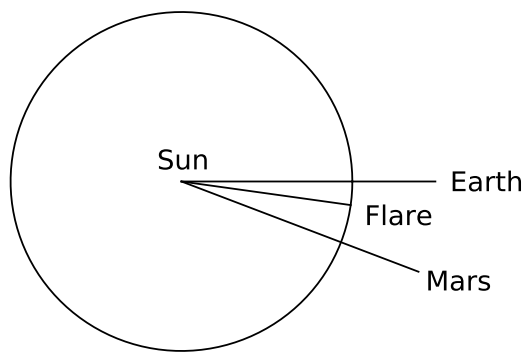}}

\caption{Relative positions of Earth, Mars and the flaring site on the Sun on 2003 October 28 (view from the heliographic north pole). The angle Earth--Sun--Flare is \ang{8}, the angle Mars--Sun--Flare is \ang{13}.} 
\label{fig:orbits}
\end{figure}

This estimate was performed by \cite{Emslie2012} with the RHESSI (\citealt{2002SoPh..210....3L}) HXR data, and the lower estimate was determined to be \(5.6 \times 10^{31}\) ergs, that is more than two orders of magnitude less than the estimate for the November 4 flare given by \cite{Kane2005}. However, it should be emphasized that the October 28 event was only partially observed by RHESSI, which apparently missed the maximum of the flare HXR impulsive phase because of the passage through the South Atlantic Anomaly (SAA; see Fig.~\ref{fig:countrates}). In contrast, this event was fully observed by the High Energy Neutron Detector (HEND), a part of the gamma-ray spectrometer GRS on board of the Mars Odyssey spacecraft (\citealt{Boynton2004}). This instrument observed the event from Mars orbit, at nearly the same angle as RHESSI (see Fig.~\ref{fig:orbits}), and it provides HXR spectral information in the range 87--1014 keV for the whole duration of the flare. This event is the most powerful of all listed by \cite{Livshits2017} in their catalog of solar flares detected by HEND in 2001--2016. The goal of this work is to use these HXR data to infer the total amount and energy of non-thermal electrons accelerated in the course of this extreme flare. This will help to refine the estimate given by \cite{Emslie2012} and compare it with the estimate given by \cite{Kane2005} for the November 4 flare. 

The paper is organized as follows. In Section~\ref{observ} the observations and the data are described. In Section~\ref{models} we discuss the models of propagation and bremsstrahlung of non-thermal electrons in the solar flare region and then apply them to infer the total amount of accelerated electrons and their energy. We also present simple estimation of amount and energetics of interplanetary energetic electrons. Discussion and conclusions are given in Sections~\ref{discuss} and \ref{conclusion} respectively. 
%%%%%%%%%%%%%%%%%%%%%%%%%%%%%%%%%%%%%%%%%%%%%%%%%%%
\section{Observations and data reduction}\label{observ}
\subsection{Mars Odyssey/HEND observations}
The HEND instrument onboard 2001 Mars Odyssey spacecraft is described by \cite{Boynton2004}. It consists of a set of $^3$He proportional counters and a scintillation block with two detectors. We used the data from the outer (CsI) scintillator. It provides the HXR count rates in 16 energy channels with the time resolution of around 20 sec. This detector was not pre-calibrated before the flight, and we used the calibration described in \cite{Livshits2017}. The energy boundaries are reliable only for channels 3--14 which cover the range of 87--1014 keV. The energy channels are shown in Table~\ref{tab:channels}.

2001 Mars Odyssey is in a polar Sun-synchronous orbit, therefore it is continuously exposed to the sunlight. However a given flare observed from near Earth may not be observed by HEND due to the relative locations of Earth, Mars and the flaring site on the solar surface. Moreover, sometimes the Sun as observed by HEND is obscured by the spacecraft, and in such cases the measurements are not reliable. We checked the observation conditions for the time interval from 2003 October 28 11:00 to 11:35 UT, when the flare took place, with the SPICE package \cite{Acton1996}. The positions of Earth, Mars and the location of the flare on the solar surface are shown schematically in Fig.~\ref{fig:orbits}. One can see that the flare was observed from Mars at about the same small angle as from Earth (Earth and Mars were located almost symmetrically with respect to the line connecting the center of the Sun and the flare site). This means that the possible effects of anisotropy of HXR emission in this event, if present (\eg, \citealt{1988ApJ...326.1017K,2012Ge&Ae..52..875K}), have the same influence on the observations of both the spacecraft and we will not consider them. The analysis of the spacecraft orientation showed that the Sun was seen directly from the position of HEND within the time of the flare.

In general, HEND data reduction was the same as in \cite{Livshits2017}. Here we briefly outline the procedure. First, the background was subtracted. The background level was estimated from the signal before the flare and linearly extrapolated up to the end of the HXR burst. The adequacy of the background level obtained was estimated visually because the flare was well pronounced in all the channels and the background showed linear behavior before and after the flare. The raw count rate and the background are shown in panel $a$ of Fig.~\ref{fig:main}. Next, the count rates were converted to the photon fluxes using the response matrices obtained from the calibration procedure and the data of the detector shape and the discriminator coefficients.

As was mentioned before, RHESSI observed the flare only partially, starting from 11:06:15 UT (see Fig.~\ref{fig:countrates}). These data can be used to check the correctness of the HEND data. We broke the time range from 11:06:15 to 11:29:30 UT, when RHESSI was observing the flare, into 20 sec intervals which is equal to the time cadence of HEND. In each of these intervals we approximated the RHESSI spectrum by a broken power law in the range 50--300 keV. Such an energy range was chosen in order to be sure that the contribution of the thermal component is negligible. On the other hand, the front detector's performance is sufficient in this range, because its effective area equals that of the rear detector at about 300 keV. Then the model photon spectra were summed within the boundaries of the HEND energy channels. These quantities were compared with the photon fluxes derived from the HEND data. 

\begin{figure} 
\centerline{\includegraphics[width=0.8\textwidth]{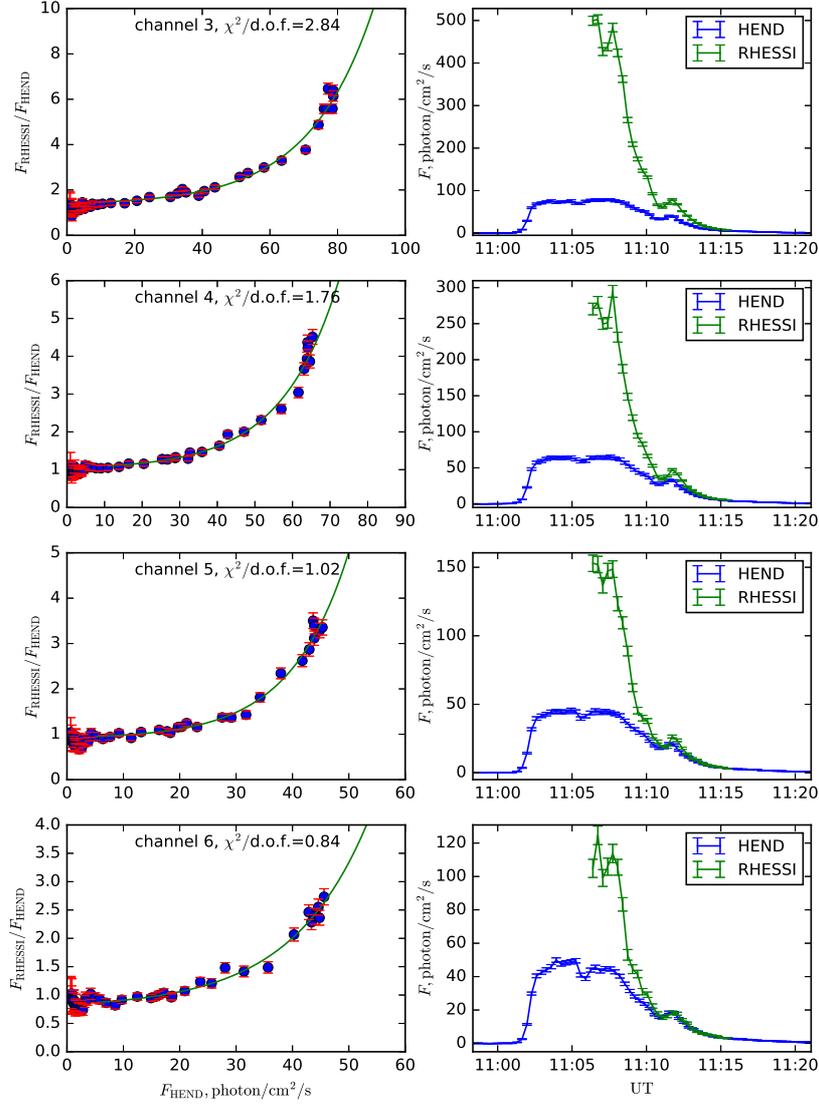}}

\caption{\textit{Left.} Ratio of the RHESSI to HEND fluxes (blue circles with errors shown by red vertical stripes) for the HEND energy channels 3, 4, 5, 6 and its fitting (green curves). \textit{Right.} Uncorrected HEND and RHESSI fluxes for the same HEND channels. The HEND fluxes are reduced to the distance from the Sun to the Earth.} 
\label{fig:saturation}
\end{figure}

This comparison revealed that HEND suffered from saturation during this powerful flare: HXR fluxes measured by HEND are lower than fluxes measured by RHESSI and the difference is larger for larger flux. This effect is caused by the HEND electronics having a finite time resolution. When the time interval between two photons detection is too short they are registered as one photon with the energy equal to the sum of the energies of the incident photons. This is known as the pile-up effect. Fig.~\ref{fig:saturation} illustrates the problem: in the left column, we show the ratio of the RHESSI flux to that of HEND for the HEND channels 3, 4, 5, 6. The curves fitted are of the form $R(x) = a + be^{x/c}$, where $a, b, c$ are the parameters fitted for each channel. For the higher energy channels the saturation curve approaches a straight line. In the right column we show the HEND and RHESSI fluxes for the same HEND channels. The HEND fluxes are reduced to the distance from the Sun to the Earth. One can see that the response of HEND continues to vary even at the high level of saturation. Therefore, we decided to use the curves shown in Fig.~\ref{fig:saturation} to correct the HEND fluxes via multiplication of the flux $F$ in the given channel by $R(F)$.

In our work we used the observations of RHESSI in the range 50--300 keV for two reasons. First, the background above 300 keV cannot be estimated in a straightforward way (see \ref{sec:rhessi}); second, the effective area of the front detectors drops down above 300 keV \cite{Smith2002}. Therefore, we applied the correction to the HEND channels 3--8, the resulting energy range being 87--285 keV. The corrected normalization coefficient and the power law index as functions of time are shown in panels $b$ and $c$ of Fig.~\ref{fig:main} respectively. In panel $c$ we also show the two power law indexes of the RHESSI photon spectra.

\begin{figure} 
\centerline{\includegraphics[scale=0.6]{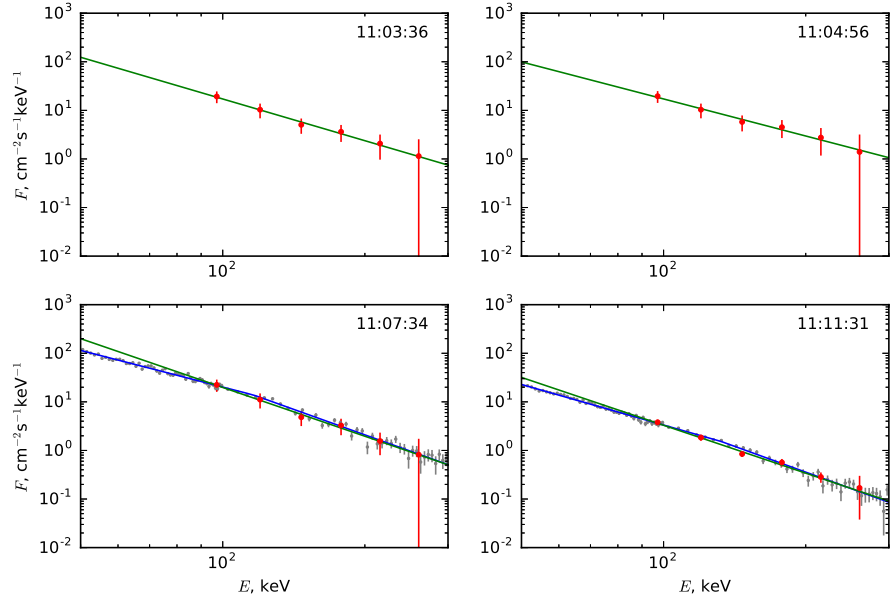}}

\caption{HEND and RHESSI (where available) spectra for four 20-second intervals shown with dark gray vertical shadings in Fig.~\ref{fig:countrates} (the start time of each interval is shown in the corner of each panel). The red dots and error bars are corrected HEND fluxes in channels 3--8. The green line represents the power law fit to the HEND data. The gray dots and error bars show RHESSI photon fluxes. The blue line shows the broken power law fit to the RHESSI spectral data.} 
\label{fig:demo_spec}
\end{figure}

The model curve of the form $I(E) = A(E/E_0)^{-\gamma}$ was fitted to the corrected flux data giving us the HXR spectrum for each time instance. In Fig.~\ref{fig:demo_spec} we show several examples of HEND and RHESSI spectra at various time instances.

The flare under consideration was observed by Konus/Wind \cite{Aptekar1995}. We compared HXR spectra in the 170-500 keV range made with HEND and Konus/Wind for several time intervals within 11:01:12-11:02:57 UT, when the Konus/Wind spectral data are available, and found consistency between these spectra within a factor of 2 (private communication with A.~Lysenko, Ioffe Institute, St.~Petersburg, Russia). This gives us additional confidence that our calibration of the HEND data is quite adequate.

\subsection{RHESSI observations}\label{sec:rhessi}
\begin{figure}
\begin{minipage}{0.45\linewidth}
\centerline{\includegraphics[scale=0.35]{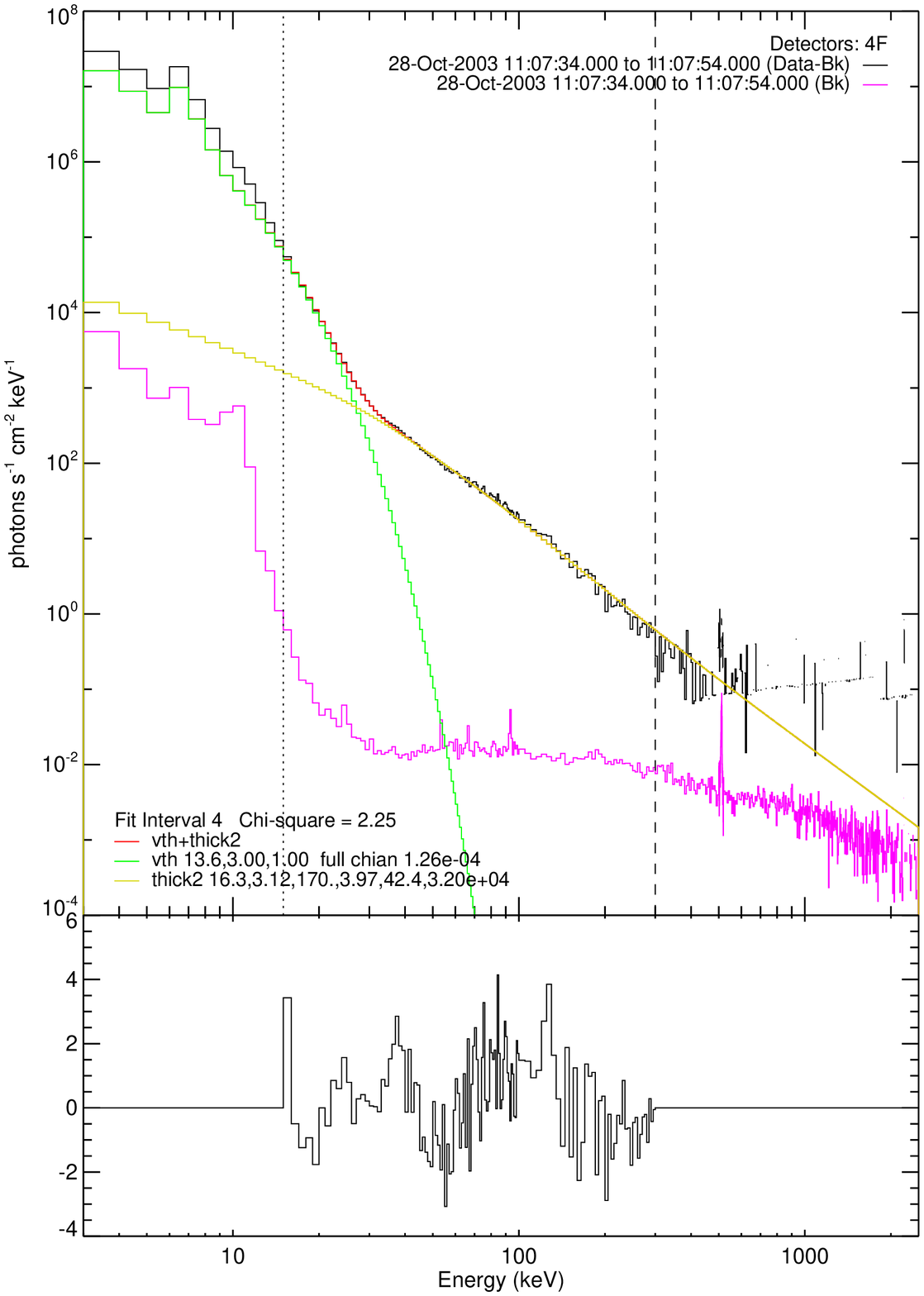}}
\end{minipage}
\begin{minipage}{0.45\linewidth}
\centerline{\includegraphics[scale=0.35]{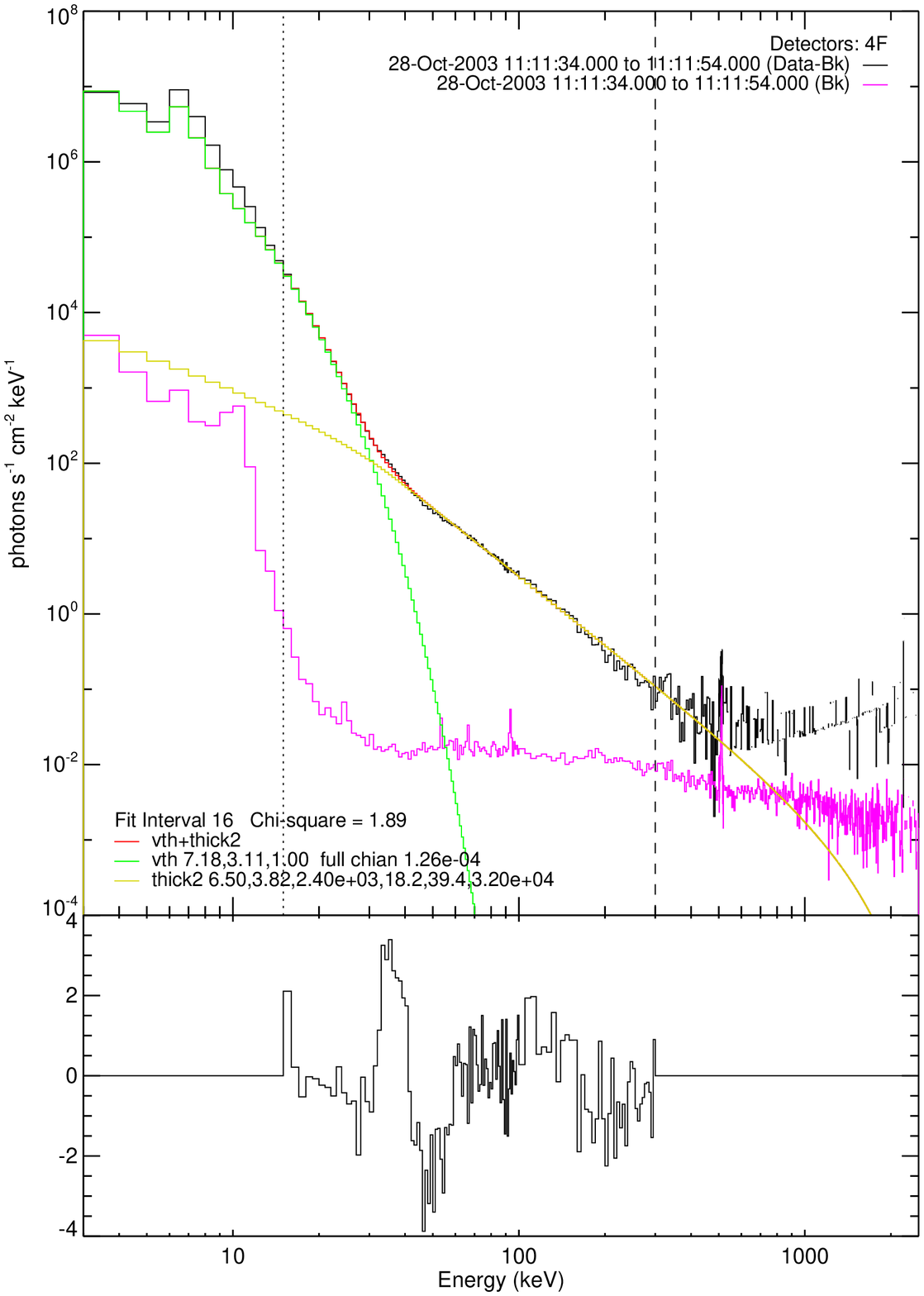}}
\end{minipage}
\caption{Examples of the RHESSI spectra fitting and residuals for the time intervals 11:07:34--11:07:54 and 11:11:34--11:11:54 UT (see bottoms panels on Fig.~\ref{fig:demo_spec}). In the bottom right plot the effect of pile-up is evident around 30-50 keV. The green and yellow curves represent the fitting functions of thermal and non-thermal components respectively.} 
\label{fig:pileup}
\end{figure}

The parameters of non-thermal electrons can be derived from the RHESSI HXR spectra using the \textit{thick2} model which is an implementation of the classic collisional thick target model in the OSPEX spectral analysis package.
We used RHESSI data from the detector 4 from 11:06:14 to 11:20:54 UT (when the RHESSI data are available and the signal is present below 300 keV). The standard pileup correction and albedo correction with the fixed up to down flux ratio equal to 1 were applied. The time interval was divided into subintervals of 20 sec length. The background was estimated in the night intervals before and after the flare (10:24:34--10:28:38 and 11:30:18--11:34:22 UT) using the 1Poly approximation. The spectrum was then approximated by the combination of an isothermal and a non-thermal components (vth + thick2) in the range from 15 to 300 keV. The lower boundary was chosen so that the isothermal approximation gave an acceptable fit. The flare under consideration was exceptionally strong and the pile-up effect could not be fully corrected by the standard RHESSI software. This is evident in Fig.~\ref{fig:pileup} where the fragments of the spectra are shown for the time intervals 11:07:34--11:07:54 and 11:11:34--11:11:54 UT together with the residuals. In the right plot, one sees a bump at 30--40 keV. We regard it as the pileup artifact which could not be fully accounted for by the standard correction performed by the hessi software. There is a possibility to use, instead of this correction, another one, namely the \textit{pileup\_mod} pseudo-function in OSPEX, used in \cite{Mann2009} for example. We tried this option and obtained good fits, but the errors of the fit parameters were too large, on the order of 100\%. The final estimations of the total number and energy of the non-thermal electrons were close in both cases, i.e. using pileup correction in HESSI and in OSPEX. Therefore we followed the first approach which gave reasonable errors.

We used the fits obtained in two ways. First, to compare the number and the energy of the non-thermal electrons derived from the other models and HEND data. Second, to obtain the temperature of the hot plasma which is a parameter of the warm target model. The data from RHESSI are available from 11:06:15. For earlier times, only GOES data are available. It is known that the temperature derived from the GOES data is systematically lower than that derived from the RHESSI data \cite{Battaglia2005,Warmuth2016}. We corrected the temperatures obtained from GOES using the relation suggested by \cite{Battaglia2005}: $T_\mr{R} = 1.13T_\mr{G} + 3.17$ MK. The evolution of the temperature with time is shown in panel $d$ of Fig.~\ref{fig:main}.

\begin{figure}
\begin{minipage}{0.25\linewidth}
\centerline{\includegraphics[scale=0.6]{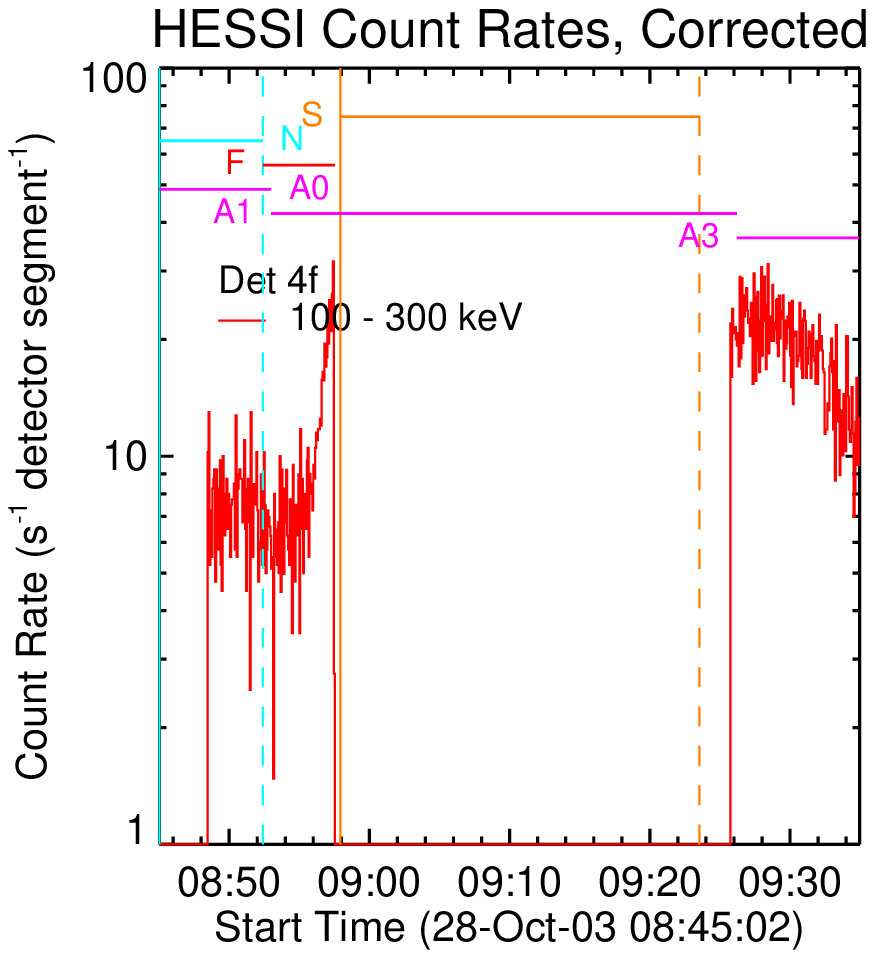}}
\end{minipage}
\hfill
\begin{minipage}{0.25\linewidth}
\centerline{\includegraphics[scale=0.6]{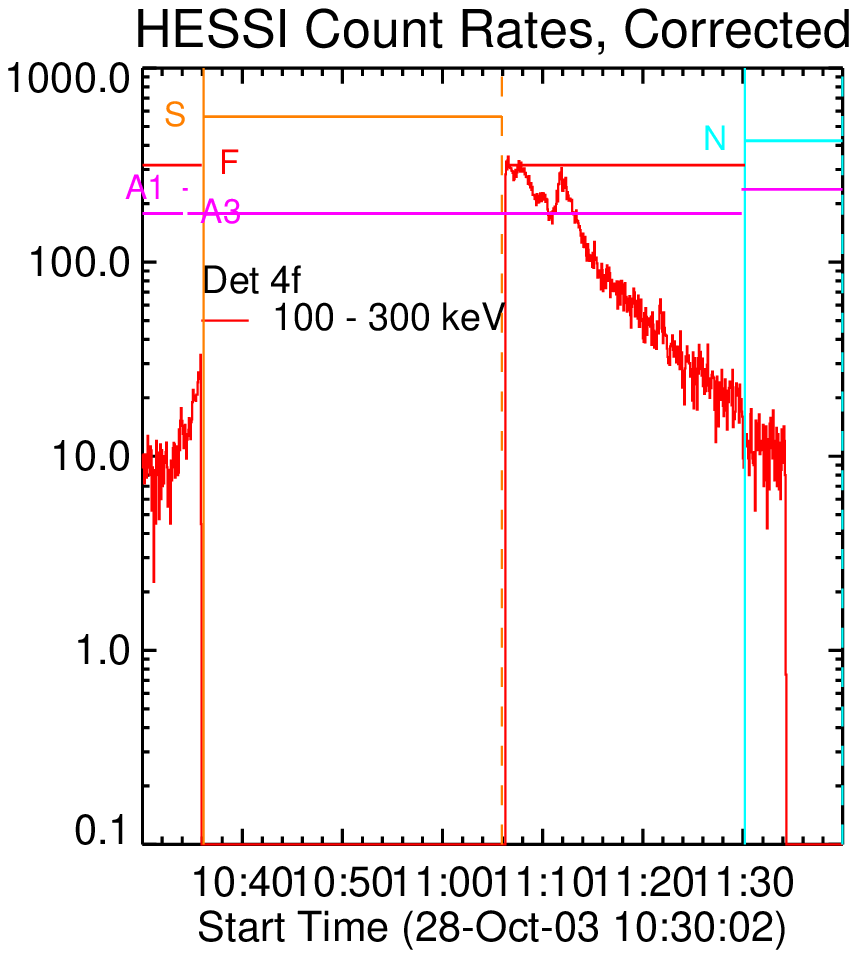}}
\end{minipage}
\hfill
\begin{minipage}{0.25\linewidth}
\centerline{\includegraphics[scale=0.6]{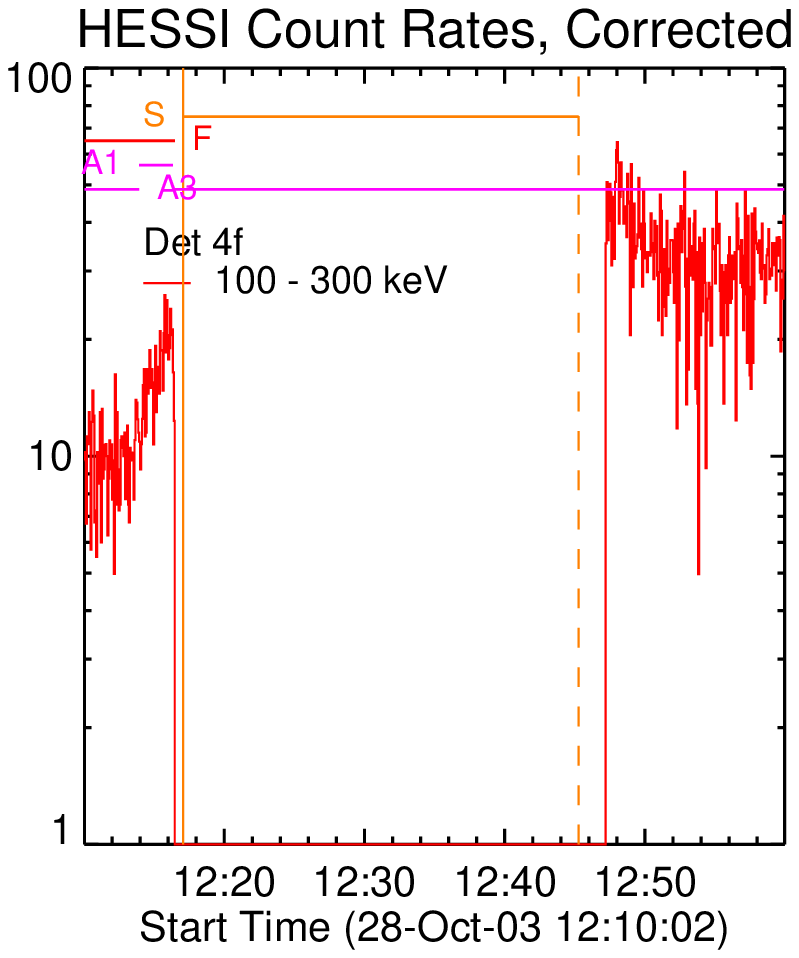}}
\end{minipage}
\caption{Quick-look plots of the corrected count rates showing the background when leaving SAA on three subsequent orbits: the one containing the impulsive phase of the flare (middle), the previous one (left) and the next one (right).}
\label{fig:saa}
\end{figure}

\begin{figure} 
\centerline{\includegraphics[scale=0.5]{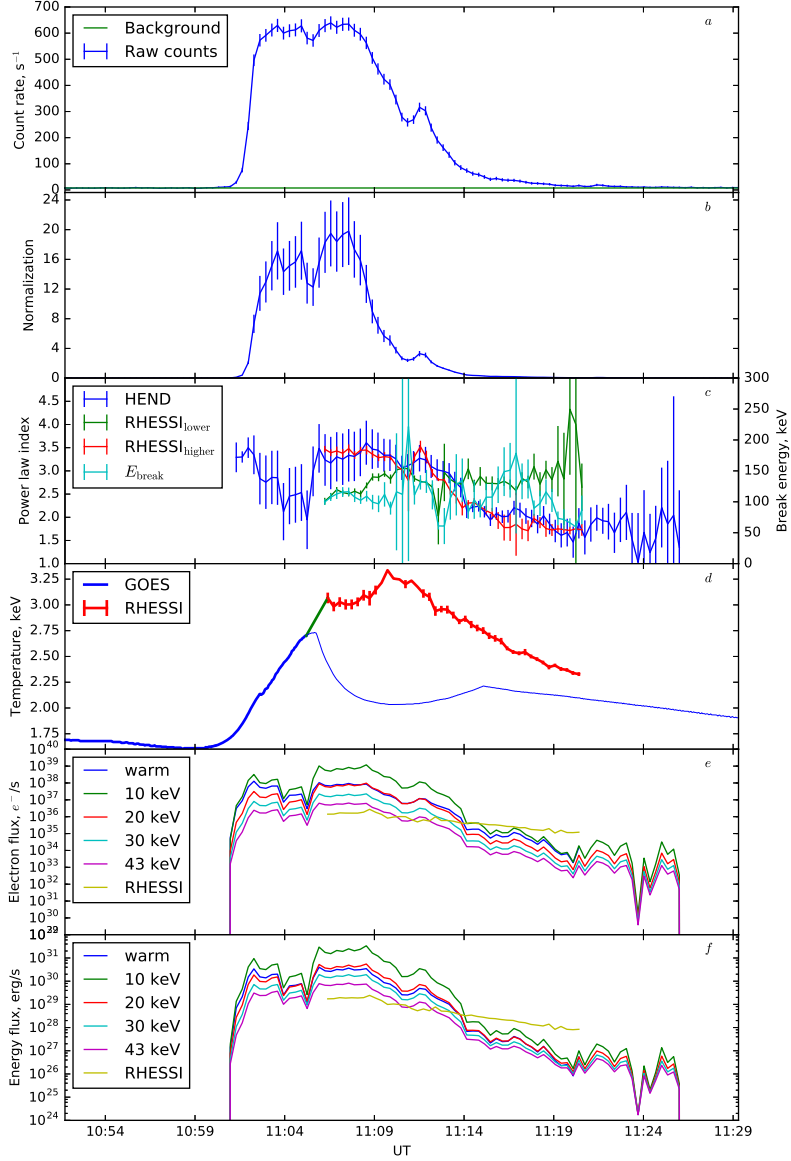}}

\caption{$a.$ Raw count rate and the background level in the HEND channel 3 (87--107 keV. $b.$ The corrected normalization factor of the HEND power law fits. $c.$ The corrected power law index of the HEND fits, the power law indices of the RHESSI photon spectra and the break energy. $d.$ The temperature of the hot flare region plasma derived from GOES (blue) and RHESSI (red). The GOES data are corrected as suggested by \cite{Battaglia2005}. The green segment is obtained via linear interpolation. The thick lines represent the data used in the analysis. The saturation of GOES is manifested as a spurious suppression. $e.$ The time variation of the total number of accelerated electrons for the models used in this work. $f.$ Same for the total energy carried by the accelerated electrons.}
\label{fig:main}
\end{figure}

A remark on the RHESSI background should be made. The observations started right after leaving SAA. Because SAA does not have sharp edges, it continues to contribute to the background several minutes after the start of the observations. In other words, the background at this time is higher than that estimated from the night intervals. In order to estimate this contribution, we compared the quick-look plots for the three time intervals: one containing the flare, one containing leaving SAA on the previous orbit and one on the next orbit. The plots for the corrected count rates (CCR) of the front detector 4 are given in Fig.~\ref{fig:saa}. The values of CCR in the 100--300 keV band when leaving SAA on the three subsequent orbits are respectively 30, 307 and 40 while the night value is 14. One sees from the quick-look plots the background corrected CCR in the 100-300 keV band should have probably been between 267 and 277 instead of 293. That is, the error introduced is about 10\% which does not affect our results severely. We note that the effect becomes more pronounced at higher energies. CCR in the 300-800 keV band when leaving SAA on the three subsequent orbits is 110, 360 and 280 while the value at night is 70 (these are the values for dets 1, 3, 4, 5, 8, 9 because single det4 provides rather low SNR in this band). So, the error can range from 15 to 260\%. However, we do not use photons above 300 keV in our analysis.

From Fig.~\ref{fig:main}~$c$, one can see that the photon spectrum follows a broken power law. During the gradual phase, the spectrum below the break steepens with time while the spectrum above the break flattens. At 11:10:34 the spectrum becomes breaking upward ($\gamma_\mr{lower} > \gamma_\mr{higher}$). In the next time interval (11:11:14-11:11:34) a secondary HXR burst begins making the spectrum breaking downward again. When this burst fades out, the difference $\gamma_\mr{lower} - \gamma_\mr{higher}$ resumes increasing and starting with 11:12:54 the spectrum is again breaking upward. The break energy also shown in Fig.~\ref{fig:main}~$c$ remains quite close to 100 keV up to 11:15:00 and does not show abrupt changes. It spikes at 11:10:54 (right before the secondary HXR peak) up to 222 keV, but it is the time when the spectrum is essentially single power law and the break energy is poorly determined, therefore the coincidence of this spurious spike on the break energy curve with the secondary HXR burst is accidental. In general, the crossover to breaking upward at the decay phase of flare HXR bursts is consistent with the observations of \cite{Dulk1992}. An example of the break upward spectral behavior during the minima between two HXR bursts can be found in \cite{Warmuth2009}.
%%%%%%%%%%%%%%%%%%%%%%%%%%%%%%%%%%%%%%%%%%%%%%%%%%%
\section{Estimation of number and energetics of non-thermal electrons }\label{models}
\subsection{Non-thermal electrons in the flare region on the Sun}
The HXR spectrum allows one to calculate the instantaneous flux of non-thermal electrons being accelerated during the flare. Different approaches are possible. First, one can accept the thick target model developed by \cite{Brown1971, Syrovatskii1972} and deduce the non-thermal electron characteristics as a solution of the inverse problem as in the articles cited. In this case one has to introduce the lower energy cutoff $E_\mr{c}$ of the electron source. Since HEND data provide no information about its value (which is usually less than 50 keV in solar flares) we perform the calculations for several cutoff energies from 10 to 43 keV. Second, one can use the warm target model by \cite{Kontar2015}, which does not introduce the cutoff energy artificially, because the electron spectrum appears to be suppressed at low energies due to the model properties. The effective cutoff energy in this model is derived from the spectral power law index and the target's temperature. Thus, in this case one needs information of the flaring plasma temperature. We utilize this approach either by using the GOES data to determine the temperature of the soft X-ray (SXR) emitting plasma when the RHESSI data were not available or using RHESSI spectral fits as described in Sec.~\ref{observ}. Finally, one can deduce the properties of the accelerated electrons from RHESSI data with the OSPEX package by forward fitting the \textit{thick2} model. This is again an invocation of the collisional thick target model, but applied purely to the RHESSI data. We follow this approach in order to compare our calculations using the HEND data. \cite{Emslie2012} estimated the total energy of non-thermal electrons in this flare using RHESSI data and thus obtained the lower limit of this quantity (see Introduction). We will refer to their results later.

In the framework of the thick target model, one needs the HXR spectrum and some guess about the electrons' low energy cutoff $E_\mr{c}$ in order to obtain the total amount of accelerated electrons and their energy. In the literature, different values of $E_\mr{c}$ are adopted, usually from 10 to 30 keV. We used RHESSI data to estimate $E_\mr{c}$ at the time it exited SAA, the value obtained is $42.9 \pm 8.3$ keV which is the highest threshold energy consistent with the observations. Hence, the estimates of the non-thermal electron number and energy are the lower estimates. Thus, we performed the calculation for four values of $E_\mr{c}$: 10, 20, 30 and 43 keV. In the following, we designate the electron energies with $E$ and the photon energies with $\varepsilon$. We assume that the HXR intensity has a power law index $\gamma$, i.e. it is proportional to $\varepsilon^{-\gamma}$. Then the source integrated electron spectrum (i.e. the spectrum of the electrons within the target) has the power law index $\gamma_\mr{el} = \gamma-1$ and the acceleration spectrum has the power law index $\delta = \gamma+1$. We used the formulas from \cite{Syrovatskii1972}:
\begin{equation}
\Phi(E > E_1, t) = 1.02 \times 10^{34} \frac{\gamma_\mr{el}^2}{E_1 \mr{B}(\gamma_\mr{el}, \frac12)} \frac{I(\varepsilon_1 < \varepsilon < \varepsilon_2)}{1 - \left(\frac{\varepsilon_1}{\varepsilon_2}\right)^{\gamma_\mr{el}}}, [\text{sec}^{-1}]
\end{equation}
for the number of the electrons and 
\begin{equation}
\mathcal{F}(E > E_1, t) = 1.02 \times 10^{34} \frac{\gamma_\mr{el} (\gamma_\mr{el} + 2)}{\mr{B}(\gamma_\mr{el}, \frac12)} \frac{I(\varepsilon_1 < \varepsilon < \varepsilon_2)}{1 - \left(\frac{\varepsilon_1}{\varepsilon_2}\right)^{\gamma_\mr{el}}}, [\text{keV sec}^{-1}]
\end{equation}
for the energy. In these formulas, $\varepsilon_1$ and $\varepsilon_2$ are the boundaries of the HXR spectrum, $I$ is the total (energy integrated) HXR flux. We set $\varepsilon_1 = E_1 = E_\mr{c}$ (\ie we extrapolate the spectrum down to $E_\mr{c}$), $\varepsilon_2 = \infty$.

The warm target model introduced by \cite{Kontar2015} is believed to avoid the difficulty of unknown $E_\mr{c}$. Due to the electron thermalization taken into account in this model, the spectrum at the low energy end becomes suppressed without introducing a cutoff. The beam cross-section integrated flux reads in this model:
\begin{equation}
AF_0(E_0) = -2K \frac{d}{dE}\left[\frac{1}{E}\left(\frac{E}{kT} - 1\right) G\left(\sqrt{\frac{E}{kT}}\right)\langle nVF\rangle(E)\right]_{E=E_0} \label{warm}
\end{equation}
where $K = 2\pi e^4 \Lambda$, $\Lambda$ is the Coulomb logarithm,
\begin{equation}
G(u) = \frac{\mr{erf}(u) - u\mr{erf}'(u)}{2u^2},
\end{equation}
$\mr{erf}$ is the error function. Further, $\langle nVF\rangle$ is the source integrated electron spectrum:
\begin{equation}
\langle nVF \rangle = \frac{\int_V n(\mathbf{r}, t) F(E, \mathbf{r}, t) dV}{\int_V n(\mathbf{r}, t) dV}
\end{equation}
which in case of the power law spectrum of the form $I(E) = A(E/E_0)^{-\gamma_\mr{el}}$ is equal to
\begin{equation}
\langle nVF \rangle = 6.66 \times 10^{47}\mr{B}(\gamma_\mr{el} - \frac32, \frac12) A (2\gamma_\mr{el} - 3)(\gamma_\mr{el} - 1)E_0^{\gamma_\mr{el}} E^{1-\gamma_\mr{el}}, [\text{cm}^{-2}\text{s}^{-1}\text{keV}^{-1}]
\end{equation}
In (\ref{warm}), $T$ is the temperature of the target. We estimated it from the GOES and RHESSI data (see Sec.~\ref{observ}). Eq.~\ref{warm} does not represent the complete warm target model, it only relates the pitch-angle averaged injection spectrum with the electron spectrum integrated over the emitting source (see \cite{Kontar2015} for details). This relation can be effectively reduced to the cold target model with $E_\mr{c} = (\gamma_\mr{el}+2)kT$, hence the number and the energy of the non-thermal electrons can be calculated using the cold target model formalism. However we calculated these quantities via direct integration of Eq.~\ref{warm} with the weight 1 or $E$. This gave results several times less than that obtained with the effective cold target approach. The reason is that the spectrum \ref{warm} differs significantly from the pure power law at the energies close to $E_\mr{c}$.

A remark should be made about the non-thermal electron parameter derivation from RHESSI data. Here, one needs only to fit the certain model to the spectrum to obtain the desired parameters. As was stated in Sec.~\ref{observ}, the spectrum of the flare in question was distorted due to the pile-up effect. For this reason, accurate determination of the electron parameters was difficult. At the same time, this is not a problem for the calculations described above because in this case one only needs the HXR spectrum (at relatively high energies) which is fit by a power law model quite reliably. However, in such calculations the problem of unknown $E_\mr{c}$ cannot be solved.

The results of our calculations are presented in Table~\ref{tab:result}. Each row corresponds to a certain model: warm target; thick target with $E_\mr{c} = 10, 20, 30, 43$ keV; and RHESSI spectral fit. The calculations in the warm and thick target models are performed for the two time intervals: 1) the interval when the RHESSI data are available and 2) the whole duration of the flare. The comparison of the results obtained with the HEND data with those obtained purely from the RHESSI data as well as the results of \cite{Emslie2012} is given in Sec.~\ref{discuss}. The time evolution of the number and energy of non-thermal electrons is shown in panels $e$ and $f$ of Fig.~\ref{fig:main} respectively for all the models used.

\subsection{Interplanetary energetic electrons}
\cite{Mewaldt05} estimated the total energy of solar energetic particles (SEP) in the range from 0.01 to 1000 MeV/nucleon to be \(\approx 5.8 \times 10^{31}\) ergs in the 2003 October 28 SEP event. \cite{Emslie2012} gave similar value of \(\approx 4.3 \times 10^{31}\) ergs. These estimates have been performed for the entire SEP population (including electrons, protons and ions) and for the total duration of the SEP event of more than 30 hours. It was inferred that energetic electrons contain no more than $18\%$ of the total estimated energy of SEP, \ie no more than $\approx 1.0 \times 10^{31}$ ergs. 

\cite{Klassen2005} showed that there were at least two populations of energetic electrons injected in the interplanetary medium during this SEP event: impulsive and gradual. Impulsive electrons were accelerated/injected in the flare impulsive phase and they were detected for around 18 min which is similar to the duration of the flare impulsive phase observed in the HXR range. Gradual electrons started to be accelerated and released approximately 20 min after the onset of the impulsive electron acceleration, and this process lasted more than 5 hours. Despite details of acceleration/injection processes in this event being unknown, the timing analysis indicates that the impulsive electrons could be accelerated during the flare impulsive phase in the flaring region, while the gradual electrons were accelerated later, and the site of their acceleration is not clear. Thus, it makes sense to compare total numbers and energetics of impulsive electrons in the interplanetary medium and non-thermal electrons in the flare region. We will also estimate the same parameters for gradual electrons.

We will use the following simplified relations to estimate total number 
\begin{equation}
N^{i,g}(E>E_{0}) \approx 2 \pi^2 L_{AU}^2 \Delta t^{i,g} \int_{E_{0}}^{\infty} I^{i,g}(E) dE 
\label{N_of_E}
\end{equation}
and total energy 
\begin{equation}
\epsilon^{i,g} (E>E_{0}) \approx 2 \pi^2 L_{AU}^2 \Delta t^{i,g} \int_{E_{0}}^{\infty} I^{i,g}(E) E dE 
\label{epsilon_of_E}
\end{equation}
of energetic electrons with energies above some threshold level $E_{0}$ injected into the interplanetary space. Here $i$ and $g$ superscripts denote impulsively and gradually injected electrons respectively. To derive these relations we made the following assumptions: (a) the electron injection functions had a symmetrical triangular shape with total duration of $\Delta t^{i}=18$ min for the impulsive injection and $\Delta t^{g}=5$ hours for the gradual injection, (b) electrons were injected homogeneously into a hemisphere with a radius ($L_{AU}$) of one astronomical unit. Since we do not know the real longitude and latitudinal distribution of particles, as well as their pitch-angular distribution, our assumptions do not appear excessively specific and are suitable for estimation with order of magnitude accuracy.

According to \cite{Klassen2005} the peak intensity spectra of impulsive and gradual electrons in this event can be fit by combinations of two power-law functions $I_{1,2}^{i,g}(E)=A_{1,2}^{i,g} \times (E/E_{1,2}^{i,g})^{-\delta_{1,2}^{i,g}}$ $[\text{cm}^{-2} \text{sec}^{-1} \text{ster}^{-1} \text{MeV}^{-1}]$, the first one is for the energies below around $E_{2}^{i,g}=66$ keV and the second one is for the energies above $E_{2}^{i,g}$. The numerical parameters of these functions for impulsive electrons are as follows: $A_{1}^{i} \approx 5 \times 10^{7}$, $A_{2}^{i} \approx 6 \times 10^{5}$, $\delta_{1}^{i} \approx 1.9$,  $\delta_{2}^{i} \approx 6.2$, and $E_{1}^{i}=8.9$ keV; and for gradual electrons: $A_{1}^{g} \approx 2 \times 10^{6}$, $A_{2}^{g} \approx 7 \times 10^{5}$, $\delta_{1}^{g} \approx 1.3$,  $\delta_{2}^{g} \approx 2.2$, and $E_{1}^{g}=27$ keV. Here we multiplied coefficients $A_{1,2}^{i,g}$ from \cite{Klassen2005} by $10^{6}$ to take the error in their paper into account (the ordinate axis dimensionality in their Fig.3 should be $[\text{cm}^{-2} \text{sec}^{-1} \text{ster}^{-1} \text{eV}^{-1}]$; see, \eg, \citealt{Simnett05,Mewaldt05}). Since $E_{0}$ is not known, we will make estimations for several different values $E_{0}^{1}=0.1$, $E_{0}^{2}=1$, $E_{0}^{3}=10$, $E_{0}^{4}=20$, $E_{0}^{5}=30$, and $E_{0}^{6}=43$ keV. $E_{0}^{1}$ corresponds to the boarder energy between the Maxwellian-distributed solar wind electrons and power-law-distributed energetic electrons (see, \eg, \citealt{lin85}). Calculating the integrals in the formulas (\ref{N_of_E}) and (\ref{epsilon_of_E}), we get the values for $N(E>E_{0}^{1,2,3,4,5,6})$ and $\epsilon (E>E_{0}^{1,2,3,4,5,6})$ summarized in Table~\ref{tab:result_sep}. 

%%%%%%%%%%%%%%%%%%%%%%%%%%%%%%%%%%%%%%%%%%%%%%%%%%%%%%%%%%%%%%
\section{Discussion} \label{discuss}
\cite{Emslie2012} estimated the energetics of 38 solar flares in various channels of energy release. The non-thermal electron energy was estimated from RHESSI HXR observations. Regarding the flare on 2003 October 28 the authors give the lower estimate $E > 5.6 \times 10^{31}$ ergs due to the fact that only a part of the flare was observed by RHESSI and they adopted the highest value of $E_\mr{c}$ which gave an acceptable spectral fit. Our lowest energy estimation derived from the HEND data within the time interval of RHESSI observations is $1.6 \times 10^{32}$ ergs for the low-energy cutoff $E_\mr{c} = 43$ keV. However, \cite{Emslie2012} point out that their estimations are the lower limits in the sense that they adopted the highest $E_\mr{c}$ which provided acceptable fit and the actual energy of non-thermal electrons can be up to one order of magnitude higher. Thus, regarding the accuracy of our estimations based on the HEND data, we argue that they are consistent within a factor of three with the analysis based on forward fitting of RHESSI spectra; this provides the grounds to consider our results valid. Our own calculations using RHESSI data are close to that of \cite{Emslie2012}. Again, as stated by \cite{Emslie2012}, it is the uncertainty of $E_\mr{c}$ which hinders accurate and reliable estimation of the non-thermal electron energy. Since our estimation in case of $E_\mr{c} = 43$ keV is quite close to that obtained from forward fitting, we argue that the value we obtained for the full time of the flare ($2.3 \times 10^{32}$ ergs) is the lower estimate of the total non-thermal electron energy released in the 2003 October 28 flare. In such case, the energy content of non-thermal electrons produced around the peak of the flare impulsive phase, missed by RHESSI, is $\approx 40 \%$ of the total energy of non-thermal electrons produced during the entire flare.

The warm target model gives significantly higher estimates of the number and energy of non-thermal electrons. The effective cutoff energy in this model is $(\gamma_\mr{el} + 2) kT$ where $\gamma_\mr{el}$ is the power law index of the HXR source integrated electron spectrum and $T$ is the target temperature. In our calculations, this effective cutoff never exceeds 12 keV, however the results of this model are closer to the results of thick target model with $E_\mr{c} = 20$ keV because the warm target model spectrum is suppressed near the effective cutoff as compared to the pure cold target spectrum. The total energy derived from the warm target model exceeds the bolometric radiated energy \cite{Kopp2005}, and, by this reason, seems unrealistic.

\cite{Kopp2005} obtained the total radiated energy of the October 28 flare between 3 and $9 \times 10^{32}$ ergs from the observations with the Total Irradiance Monitor. This is consistent with the estimate of the free magnetic energy in the flare region by \cite{Emslie2012}. It is clear that the free magnetic energy should be larger than the energy of non-thermal electrons, and the latter should be a fraction of the total radiated energy of the flare. This indicates that the total energy of non-thermal electrons should not be more than a few $10^{32}$ ergs. This is consistent with our estimations for $E_\mr{c}=30$ and $E_\mr{c}=43$ keV within the thick target model. On the other hand, the estimations of the bolometric irradiance and the free magnetic energy rule out our results obtained for the warm target model and the cold target model with $E_\mr{c} = 20$ keV or lower. Note that the warm target model requires the information on the target's length and density which is not available in this case. Therefore what we used is only an approximation to this model which allows to treat it like the cold target model with an effective low-energy cutoff. Such an approach is used e.g. by \cite{Aschwanden2016} but their results also seem to be significantly overestimated. The constraint set by \cite{Kopp2005} also indicates that, within the cold target model, the lower energy cutoff is greater than 20 keV and in fact, can exceed 40 keV, although in our analysis this value is an upper limit.

It is interesting to compare our results with the results for another giant flare on 2003 November 4, whose X-ray class might be even higher than that of the 28 October flare (\eg, \citealt{Kiplinger04,Brodrick05}). \cite{Kane2005} give the estimate of $\sim 3 \times 10^{41}$ electrons above 20 keV and the energy between $4 \times 10^{33}$ and $3 \times 10^{34}$ ergs that is several times larger than our result for $E_\mr{c}=20$ keV (\ie $1.3 \times 10^{33}$ ergs). It should be noted that the threshold energy of 20 keV is only assumed by \cite{Kane2005} and not derived from the observations. At the same time, the estimate of the bolometric energy of this flare provided by \cite{Emslie2012} is $4 \times 10^{32}$ erg. It means that $E_\mr{c}$ was significantly greater than 20 keV which we see in the flare on 28 October 2003. Unfortunately, HEND did not observe the November 4 flare since it was in the safe mode because of strong bombardment by SEP caused by the October 28--29 flares.

Finally, we obtained the estimates of the total number and energetics of the impulsive and gradual interplanetary non-thermal electrons in the 2003 October 28-29 SEP event based on the peak intensity spectra measured in the Sun-Earth Lagrangian point $\text{L}_{1}$ and presented by \cite{Klassen2005}. Our estimate of the energetics of the gradual population of the interplanetary electrons is about five times less than the estimate of \citealt{Mewaldt05}. This difference is understandable, since we integrated over 5 h, not 30 h, as in the cited work. Also, we did not take into account the longitude dependence of the intensity of interplanetary particles. Our estimates show that the amount and energetics of the impulsive interplanetary electrons is one or two orders of magnitude less than the following gradual electrons and negligible (less than $0.1\%$) compared to the corresponding values of the HXR-emitting non-thermal electrons with the same low-energy cutoff. This is a well-known situation (\eg, \citealt{Lin71,Krucker07}). It does not contradict the possibility that the same acceleration process in the flare impulsive phase is responsible for both the HXR-emitting not-thermal electrons and population of impulsive electrons escaped into the interplanetary medium from the solar corona. Why only a small fraction of the accelerated electrons escapes into the interplanetary space is still an open question. 

\cite{Aulanier2013} used numerical MHD simulations and historical data on sunspots to derive the maximum possible energy of a solar flare. Taking into account the limitations and uncertainties of their method they put the upper limit of $\sim 6 \times 10^{33}$ ergs. At the same time, flares with energies of up to $\sim 10^{36}$ ergs are observed on the Sun-type stars. Even our largest energy estimation resulting from the thick target model and fixed $E_\mr{c}=10$ keV gives an energy ($\approx 5.7 \times 10^{33}$ ergs) slightly lower than the limit of \cite{Aulanier2013}, which means that purely HXR observations (and non-thermal electron energetics derived from them) of this particular flare cannot remove the energetic gap between the strongest solar flares and flares on the Sun-type stars. Probably some other dynamo mechanism  should operate on stars with superflares (see, \eg, \citealt{Katsova2018, Brandenburg2018}), but this topic is outside the scope of the present work.
	
%%%%%%%%%%%%%%%%%%%%%%%%%%%%%%%%%%%%%%%%%%%%%%%%%%%%%%%%%%%%%
\section{Conclusion} \label{conclusion}
In this work we estimated the total amount and energy of the non-thermal electrons accelerated in the giant ($>$X17) solar flare on 2003 October 28. We used the HXR observations by Mars Odyssey/HEND which covered the full time of the flare and the viewing angle was practically the same as from the Earth. We obtained estimates for the thick target model with several values of the lower cutoff energy and for the warm target model. The results were compared with those derived from the RHESSI observations of a portion of this flare. We conclude that the result consistent with the RHESSI observations (and with other energy estimations made by other authors) is that of the thick target model with $E_{c} = 43$ keV, the results being $1.7 \times 10^{39}$ electrons and $2.3 \times 10^{32}$ ergs. The estimate obtained from the warm target model is close to that of the thick target model with $E_{c} = 20$ keV, namely $2.9 \times 10^{40}$ electrons and $1.1 \times 10^{33}$ ergs. Previously, it was not known to what extent the non-thermal energy derived for this flare from the RHESSI data is underestimated due to the partial time coverage. Our results indicate that RHESSI 'missed' $\approx 40$\% of this energy.

We also estimated the number and energy of the interplanetary energetic electrons originating from this event. Our estimates for the low-energy cutoff of $E_{0}=10$ keV are $1.8 \times 10^{36}$ electrons and $6.8 \times 10^{28}$ ergs for the impulsive injection, and $1.1 \times 10^{37}$ electrons and $2.3 \times 10^{30}$ ergs for the gradual injection in this SEP event. These values are negligible comparing with the values for the HXR-generating electrons in the solar atmosphere. This is in agreement with the estimates made for other SEP events.

Our observations with HEND of the whole flare complement those made with RHESSI by \cite{Emslie2012} for a part of the flare. The result we obtained can be considered as exceptional for the observations of the Sun in hard X-rays, but it remains within the limits of possible energies according to the theoretical estimations.
%%%%%%%%%%%%%%%%%%%%%%%%%%%%%%%%%%%%%%%%%%%%%%%%%%%%%%%%%%%%%
\section*{Acknowledgements}
We are grateful to the RHESSI, GOES, and INTEGRAL teams for the available data used in this work. We also thank Dr. A.B.~Struminsky for a number of useful criticisms, A.L.~Lysenko for the help with the Konus/Wind data, and Dr. C.R.~Goddard for help with correcting the language. We express our gratitude to the anonymous reviewers for a number of constructive comments. This work is supported by the Russian Science Foundation under grant 17-72-20134.    
%%%%%%%%%%%%%%%%%%%%%%%%%%%%%%%%%%%%%%%%%%%%%%%%%%%%%%%%%%%%%

%%%%%%%%%%%%%%%%%%%%%%%%%%%%%%%%%%%%%%%%%%%%%%%%%%%%%%%%%%%%%
%\section{Bibliography styles}
\section*{References}
\bibliography{october03_rev1_clean}
%%%%%%%%%%%%%%%%%%%%%%%%%%%%%%%%%%%%%%%%%%%%%%%%%%%%%%%%%%%%%

%%%%%%%%%%%%%%%%%%%%%%%%%%%%%%%%%%%%%%%%%%%%%%%%%%%%%%%%%%%%%
%%% Figure 1
\newpage
%%%%%%%%%%%%%%%%%%%%%%%%%%%%%%%%%%%%%%%%%%%%%%%%%%%%%%%%%%%%%

%%% Figure 2
\newpage
%%%%%%%%%%%%%%%%%%%%%%%%%%%%%%%%%%%%%%%%%%%%%%%%%%%%%%%%%%%%%

%%% Figure 3
\newpage
%%%%%%%%%%%%%%%%%%%%%%%%%%%%%%%%%%%%%%%%%%%%%%%%%%%%%%%%%%%%%

%%% Figure 4
\newpage
%%%%%%%%%%%%%%%%%%%%%%%%%%%%%%%%%%%%%%%%%%%%%%%%%%%%%%%%%%%%%

%%% Figure 5
\newpage
%%%%%%%%%%%%%%%%%%%%%%%%%%%%%%%%%%%%%%%%%%%%%%%%%%%%%%%%%%%%%

%%% Figure 6
\newpage
%%%%%%%%%%%%%%%%%%%%%%%%%%%%%%%%%%%%%%%%%%%%%%%%%%%%%%%%%%%%%

%%% Figure 7
\newpage
%%%%%%%%%%%%%%%%%%%%%%%%%%%%%%%%%%%%%%%%%%%%%%%%%%%%%%%%%%%%%
\newpage
\begin{table}
\caption{HEND energy channels used in the work.}
\label{tab:channels}
\begin{tabular}{c|c|c}
\hline
Channel & Lower boundary, keV & Upper boundary, keV \\
\hline
3 & 86.568 & 107.683 \\
4 & 107.683 & 132.256 \\
5 & 132.256 & 161.056 \\
6 & 161.056 & 195.079 \\
7 & 195.079 & 235.632 \\
8 & 235.632 & 284.493 \\
9 & 284.493 & 344.09 \\
10 & 344.09 & 417.893 \\
11 & 417.893 & 510.993 \\
12 & 510.993 & 631.240 \\
13 & 631.240 & 791.506 \\
14 & 791.506 & 1014.72 \\
\hline
\end{tabular}
\end{table}

\begin{table}
\caption{Total electron number and energy calculated using warm target model and cold target model with various cutoff energies adopted. In the columns marked "RHESSI" the results are given for the time interval when RHESSI was observing the flare. In columns marked "full time" the results are given for the whole time span of the flare.} \label{tab:result}
\begin{tabular}{c|c|c|c|c}
\hline
\multirow{2}{*}{Model} &
\multicolumn{2}{c|}{with RHESSI} &
\multicolumn{2}{c}{full time} \\
\cline{2-5}
 & electrons & energy, ergs & electrons & energy, ergs \\
\hline
%warm target & $2.9 \times 10^{41}$ & $7.3 \times 10^{33}$ & $3.9 \times 10^{41}$ & $1.0 \times 10^{34}$ \\
warm target & $1.7 \times 10^{40}$ & $6.6 \times 10^{32}$ & $2.9 \times 10^{40}$ & $1.1 \times 10^{33}$ \\
10 keV & $1.6 \times 10^{41}$ & $4.7 \times 10^{33}$ & $2.1 \times 10^{41}$ & $6.2 \times 10^{33}$ \\
20 keV & $1.6 \times 10^{40}$ & $9.3 \times 10^{32}$ & $2.1 \times 10^{40}$ & $1.3 \times 10^{33}$ \\
30 keV & $4.1 \times 10^{39}$ & $3.6 \times 10^{32}$ & $5.5 \times 10^{39}$ & $5.1 \times 10^{32}$ \\
43 keV & $1.2 \times 10^{39}$ & $1.6 \times 10^{32}$ & $1.7 \times 10^{39}$ & $2.3 \times 10^{32}$ \\
RHESSI & $6.3 \times 10^{38}$ & $6.3 \times 10^{31}$ & --- & --- \\
\hline
\end{tabular}
\end{table}

\begin{table}
\caption{Total number and energy of interplanetary energetic electrons with energies higher than $E_{0}$.} \label{tab:result_sep}
\begin{tabular}{c|c|c|c|c}
\hline
\multirow{2}{*}{$E_{0}, \text{keV}$} &
\multicolumn{2}{c|}{Impulsive} &
\multicolumn{2}{c}{Gradual} \\
\cline{2-5}
 & electrons & energy, ergs & electrons & energy, ergs \\
\hline
    0.1& $1.3 \times 10^{38}$ & $1.8 \times 10^{29}$ & $6.9 \times 10^{37}$ & $2.4 \times 10^{30}$  \\
1& $1.7 \times 10^{37}$ & $1.3 \times 10^{29}$ & $3.1 \times 10^{37}$ & $2.4 \times 10^{30}$  \\
  10& $1.8 \times 10^{36}$ & $6.8 \times 10^{28}$ & $1.1 \times 10^{37}$ & $2.3 \times 10^{30}$  \\
20& $7.9 \times 10^{35}$ & $4.6 \times 10^{28}$ & $7.8 \times 10^{36}$ & $2.2 \times 10^{30}$  \\
30& $4.4 \times 10^{35}$ & $3.3 \times 10^{28}$ & $6.0 \times 10^{36}$ & $2.1 \times 10^{30}$  \\
43& $2.2 \times 10^{35}$ & $2.0 \times 10^{28}$ & $4.6 \times 10^{36}$ & $2.1 \times 10^{30}$  \\
\hline
\end{tabular}
\end{table}

%%%%%%%%%%%%%%%%%%%%%%%%%%%%%%%%%%%%%%%%%%%%%%%%%%%%%%%%%%%%%
% %% Table 1
% \newpage
% \begin{table}
% \caption{Geometrical characteristics of the reconstructed MFRs ($\mathcal{T}_{\mathrm{MFR}}$ --- maximal twist number, $L_{\mathrm{MFR}}$ --- length, and $H_{\mathrm{MFR}}$ --- height), characteristics of the HXR pulsations ($n_{p}$ --- number of pulsations, $\left\langle P \right\rangle$ --- average time difference between neighboring pulsations), and two different classifications of the flares studied.}
% \begin{tabular}{|c|c|c|c|c|c|c|c|}
% \hline
% 1     &   2      &      3     &   4    &   5 & 6 & 7 & 8  \\
% \hline
% Flare &   $\mathcal{T}_{\mathrm{MFR}}$ & $L_{\mathrm{MFR}}$,~Mm & $H_{\mathrm{MFR}}$,~Mm & $n_{p}$ & $\left\langle P \right\rangle$,~s & group & case \\
% \hline
% 15-Feb-2011 & $\approx 1$ & 44 & 14 & 35 & 16 & 2 & a\\
% 07-Jun-2011 & $\approx 1$ & 46 & 13 & 36 & 26 & 1 & b \\
% 06-Sep-2011 & $\approx 1$ & 25 & 9  & 20 & 21 & 1 & b \\
% 18-Apr-2014 & $\approx 1$ & 46 & 31 & 23 & 24 & 2 & b \\ 
% 22-Oct-2014 & $\approx 1$ & 64 & 20 & 10 & 21 & 1 & b \\
% 24-Oct-2014 & $\approx 1$ & 134 & 31 & 34 & 17 & 2 & a \\
% 09-Nov-2014 & $\approx 1$ & 24 & 6  & 8  & 20 & 1 & a \\
% \hline
% \end{tabular}
% \label{tab1}
% \end{table}
% %%%%%%%%%%%%%%%%%%%%%%%%%%%%%%%%%%%%%%%%%%%%%%%%%%%%%%%%%%%%%
 
\end{document}